# Programmable ultra-broadband photonic chaos platform enabled by microwave-chaos-driven electro-optic frequency combs

Shiyu Shi[1,2,4,†], Yiqun Zhang[1,2,†], Mengjie Zhou[3,†], Mingfeng Xu[1,2,4,*], Xianglei Yan[1,2], Qiang Chen[1,2,4], Yunxia Yang[1,2], Yinghui Guo[1,2,4], Mingbo Pu[1,2,4,*], and Xiangang Luo[1,2,4,*]

[1]State Key Laboratory of Optical Field Manipulation Science and Technology, Institute of Optics and Electronics, Chinese Academy of Sciences, Chengdu 610209, China.

[2]Research Center on Vector Optical Fields, Institute of Optics and Electronics, Chinese Academy of Sciences, Chengdu 610209, China.

[3]Tianfu Xinglong Lake Laboratory, Chengdu 610299, China.

[4]College of Materials Science and Opto-Electronic Technology, University of Chinese Academy of Sciences, Beijing 100049, China.

†These authors contributed equally to this work.

*Email: xumf@ioe.ac.cn, pmb@ioe.ac.cn, lxg@ioe.ac.cn

Optical chaos holds great promise for secure communication, LiDAR, and reinforcement learning. However, its scalability has long been constrained by an intrinsic trade-off between bandwidth and the number of parallel chaotic channels. Here, we introduce a programmable "chaos-on-comb" architecture that overcomes this limitation using standard electro-optic components. By heterodyning a delayed-feedback chaotic laser with a continuous-wave reference, a broadband chaotic microwave signal is generated to simultaneously drive a cascaded electro-optic comb, imprinting chaotic dynamics across all comb lines and merging them into an ultra-broadband chaotic continuum. Then, incorporating spectrum slicing enables flexible extraction of parallel chaotic channels with preserved statistical independence and per-channel programmability. As a result, we demonstrate a single-channel ultra-broadband optical chaos with an effective bandwidth of 543.8 GHz, and a broadband terahertz noise source with an excess noise ratio of 52.99 ± 2.85 dB to validate its flatness. Furthermore, we employ the uncorrelated parallel chaos for ultrafast photonic decision-making in a 256-armed bandit problem, achieving a favourable power-law scaling exponent of 0.86. Our work paves the way toward programmable, reconfigurable, and application-ready photonic chaos systems.

**Introduction**

Chaotic systems exhibit intrinsically complex behavior due to their extreme sensitivity to initial conditions and aperiodic dynamics, which has long motivated sustained research across mathematics, physics, and engineering[1,2]. In particular, optical chaos, featuring ultrafast temporal dynamics and broad spectral bandwidth, offers distinct advantages over its electronic counterparts. In optical cryptography, laser-induced broadband chaotic carriers can conceal low-amplitude information within high-entropy waveforms, enabling physical-layer secure transmission[3-6]. In computational science, applications such as Monte Carlo simulations, stochastic modeling, and reinforcement learning rely on high-quality random number generation, for which chaos offers a critical entropy source for algorithmic and infrastructure-level operations[7-10]. For millimeter-wave noise generation, optical chaos with flat spectral profiles is recognized as a leading candidate for broadband entropy[11,12]. More recently, parallel optical chaos has enabled advances in sensing, particularly in light detection and ranging (LiDAR) systems, where it facilitates high-precision, interference-immune ranging[13-15]. Critically, the effectiveness of such applications hinges on the quality of the chaotic light source. As data demands accelerate, expanding both spectral bandwidth and parallel channel count is crucial to scaling information throughput and supporting multi-task collaborative processing.

External cavity perturbations, such as optical feedback or injection, remain the most established methods for generating chaotic lasers, yet their spectral bandwidth and flatness are ultimately constrained by the intrinsic relaxation oscillation of semiconductor gain media[16,17]. Nonlinear spectral broadening techniques can extend single-channel chaos beyond 50 GHz[18], but fail to yield a large number of uncorrelated channels. Parallel chaos generation has emerged as a promising alternative to overcome this scalability bottleneck. However, the existing parallelization strategies encountered significant limitations. Multimode lasers, which excite multiple lasing modes to generate parallel chaotic channels, suffer from strong intermodal coupling and competition, resulting in high channel correlation and limited tunability[19]. Alternatively, chaotic laser arrays or globally coupled laser networks offer spatially parallel chaos generation, but their scalability is hindered by considerable footprint and system complexity, posing challenges for miniaturization and large-scale integration[20-22]. Moreover, Microcomb-based chaotic sources operated in the modulation instability regime have demonstrated over 100 channels[23-25]. Nonetheless, due to the dispersion and nonlinear stability of microcavities, the maximum effective bandwidth is generally below 7 GHz, and their spectral properties are largely fixed by device geometry, limiting reconfigurability. To date, no reported approach has simultaneously delivered ultra-broadband, low-correlation, and highly programmable parallel chaos suitable for scalable, application-ready deployment.

Here, we overcome the longstanding trade-off between chaotic bandwidth and parallelism by introducing a novel "chaos-on-comb" architecture, enabled by a microwave-chaos-driven electro-optic frequency comb. A 32.5 GHz broadband microwave chaotic signal, generated by heterodyning a delayed-feedback chaotic laser with a continuous-wave (CW) reference, drives a common cascaded electro-optic frequency comb. Chaotic fluctuations in instantaneous frequency and amplitude are imprinted onto each comb tooth, and phase modulation induces spectral broadening and overlapping, transforming the periodic comb into a continuum-like optical chaotic source. We achieve a single-channel ultra-broadband optical chaos with an effective bandwidth of 543.8 GHz and demonstrate arbitrary, independently programmable control of parallel chaotic channels via spectrum slicing. We validate its flat spectral properties by demonstrating a 0-500 GHz noise source with an excess noise ratio (ENR) of $52.99 \pm 2.85$ dB. Additionally, we apply the uncorrelated parallel chaos to a scalable multi-armed bandit (MAB) problem in reinforcement learning, achieving superior decision speed and scalability. This platform overcomes the relaxation oscillation limit in conventional chaotic lasers, resolves the bandwidth-parallelism bottleneck in existing systems, and offers a reconfigurable foundation for next-generation applications in secure communications, AI photonics, and sensing.

## Results

**Principle of the programmable ultra-broadband photonic chaos generation**

Figure 1a illustrates the principle of the proposed "chaos-on-comb" architecture for generating the programmable ultra-broadband photonic chaos. The system begins with an optical chaotic seed at center frequency $\omega_0$, which is heterodyned with a tunable CW reference at $\omega_0+\Delta\omega$ to produce a broadband microwave chaotic signal centered at $\Delta\omega$. This signal retains the original chaotic dynamics and is used to simultaneously drive a cascaded electro-optic comb consisting of one intensity modulator (IM) and two phase modulators (PMs) in series, imprinting chaos onto each comb line and forming the basis of the ultra-broadband optical chaotic source. Figure 1b and 1c depict the evolution of the optical and radio-frequency spectra during the modulation process. The microwave chaotic signal first drives an IM, converting one CW laser into an initial optical frequency comb with symmetric high-order sidebands spaced by $\Delta\omega$. Unlike discrete spectral lines produced by monochromatic driving, the broadband microwave chaotic input results in a continuous, randomly modulated spectral envelope with non-periodic fluctuations. This signal then passes through two cascaded PMs, each driven by the same microwave chaotic waveform. The first phase modulation introduces additional sidebands, and their strong spectral overlap leads to aliasing and smoothing across the spectrum, effectively flattening the envelope. A second phase modulation stage further enhances this broadening, ultimately forming a wide, continuum-like chaotic optical spectrum. The resulting signal is transmitted through single-mode fiber (SMF), where chromatic dispersion converts the time-varying phase fluctuations into measurable intensity dynamics. This dispersive mapping not only enables direct detection but also increases the system's dynamical complexity, resulting in a high-dimensional, ultra-broadband chaotic optical signal (see Supplementary Note 1 for detailed theoretical model of the proposed scheme).

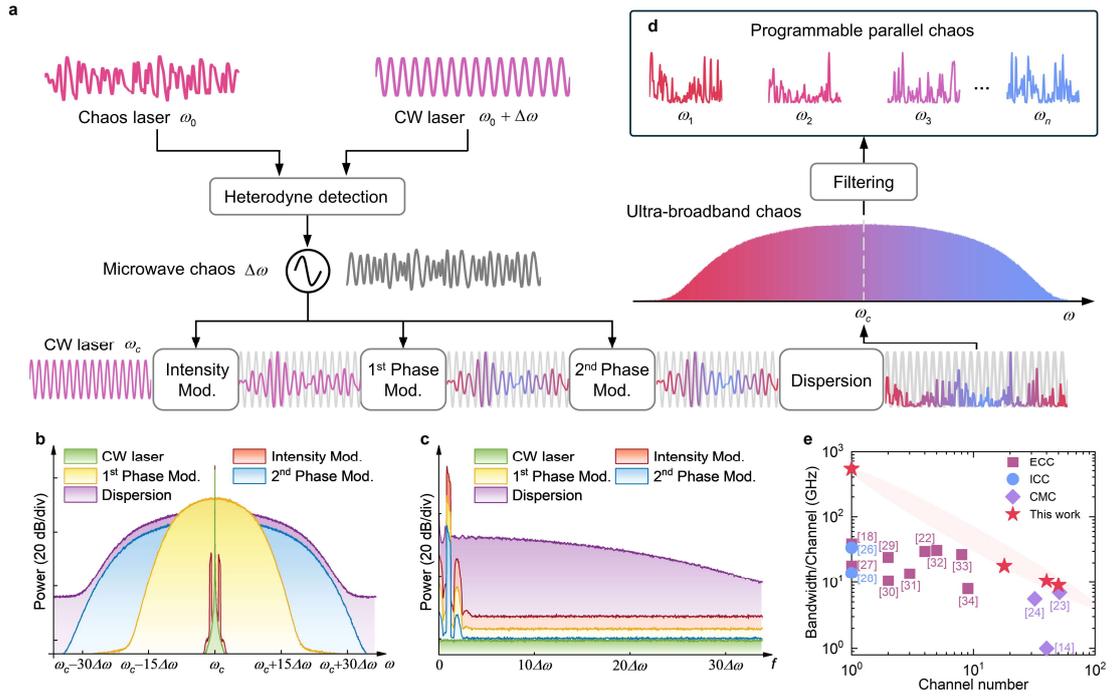

**Fig. 1 Principle and performance of the programmable ultra-broadband photonic chaos generation scheme. a**, Schematic of the proposed "chaos-on-comb" architecture. An optical chaotic seed (centered at $\omega_0$) and a CW reference (centered at $\omega_0+\Delta\omega$) are heterodyned to generate a broadband chaotic microwave signal (centered at $\Delta\omega$), which simultaneously drives a cascaded electro-optic comb consisting of one intensity modulator and two phase modulators. Phase-to-intensity conversion via dispersion produces the ultra-broadband optical chaotic source. Finally, tunable multi-channel chaotic outputs are enabled by a programmable optical filter. **b-c**, Evolution of the optical (b) and radio-frequency (c) spectra through the modulation process shown in **a**. **d**, Examples of reconfigurable parallel chaotic outputs achieved by adjusting the filter center wavelengths and bandwidths. **e**, Performance comparison between our work and prior approaches in terms of single-channel bandwidth and number of parallel chaotic channels. ECC, external-cavity chaos; ICC, internal-cavity chaos; CMC, chaotic microcomb.

The resulting ultra-broadband chaotic signal is dynamically spectrally sliced using a programmable waveshaper, enabling parallel generation of multiple low-correlation chaotic optical channels (Fig. 1d). By tuning parameters such as centre wavelength and filter bandwidth, the system can be reconfigured on demand to produce arbitrary multi-channel outputs with user-defined bandwidths, supporting flexible and application-specific deployment. As illustrated in Fig. 1e, the proposed architecture simultaneously outperforms existing approaches in both single-channel bandwidth and parallel chaotic output flexibility. We demonstrate an ultra-broadband photonic chaos source with an effective bandwidth of up to 543.8 GHz, exceeding the current state-of-the-art by an order of magnitude. Leveraging spectral slicing, this continuum enables scalable parallelization, with experimental results confirming the generation of 50 independent chaotic channels, each maintaining a bandwidth of ~9 GHz under uniform filtering conditions. This parallel performance rivals that of chaotic microcomb-based systems while offering enhanced bandwidth and reconfigurability (see more details in Supplementary Note 2).

**Characteristics of programmable ultra-broadband photonic chaos**

Figure 2a presents the experimental setup of the programmable ultra-broadband photonic chaos source (See "Methods" for more details). We first characterize the generated ultra-broadband chaotic signal. As shown in Fig. 2b, the optical spectrum exhibits a broad, flat profile with a 3-dB bandwidth of 746.25 GHz, exceeding the current state-of-the-art by a factor of seven[35]. Due to the limited electronic detection bandwidth, direct measurement across the entire spectrum was infeasible. To address this, we constructed a numerical simulation in VPIphotonics software[36], with system parameters closely matching the experimental configuration (Supplementary Table 1). The simulated spectrum (dashed line in Fig. 2b) shows excellent agreement with the experimental results, validating both the spectral span and flatness of the generated optical chaos. Figure 2c shows the time series of the generated ultra-broadband chaotic signal, with strong stochastic amplitude fluctuations. The corresponding radio-frequency spectrum in Fig. 2d reveals a flat power spectral density extending from direct current (DC), with no discernible relaxation oscillation peaks. The effective bandwidth, defined as the frequency range containing 80% of the total signal energy[37], reaches 543.8 GHz, exceeding the current international benchmarks by an order of magnitude. Figure 2e shows the auto-correlation function (ACF) of the ultra-broadband chaotic signal, exhibiting a sharp delta-like profile with no discernible time-delay signature. The full-width at half-maximum (FWHM) of the ACF is 0.89 ps, corresponding well to the spectral bandwidth.

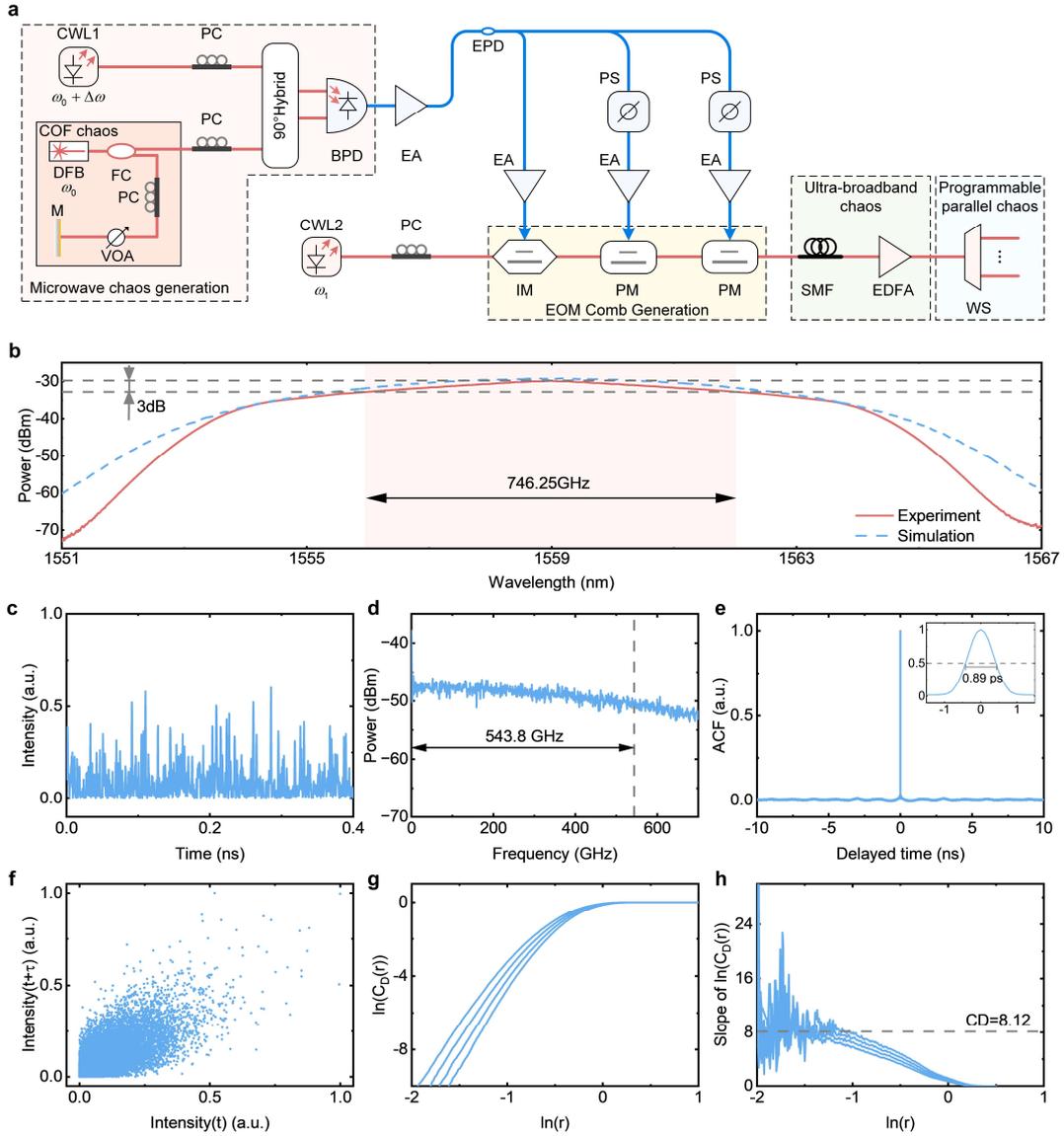

**Fig. 2 Characteristics of the ultra-broadband optical chaotic signal. a**, Experimental setup. CWL, continuous-wave laser; DFB, distributed-feedback laser; FC, fiber coupler; PC, polarization controller; VOA, variable optical attenuator; M, mirror; BPD, balanced photodetector; EA, electrical amplifier; EPD, electrical power divider; PS, phase shifter; IM, intensity modulator; PM, phase modulator; SMF, single-mode fiber; EDFA, erbium-doped fiber amplifier; WS, waveshaper; COF, conventional optical feedback. **b**, Simulated and experimental optical spectra of the generated ultra-broadband chaotic signal. **c-f**, Simulation results of the time series (**c**), radio-frequency spectrum (**d**), autocorrelation function (**e**), and phase diagram (**f**) of the ultra-broadband chaotic signal. **g**, Correlation integral $C_D(r)$ versus radius $r$. **h**, The slope of the curve in **g** is convergent to a correlation dimension of CD ≈ 8.12. The inset in **e** shows the zoom-in view of ACF, with a full width at half maximum of 0.89 ps

Figure 2f presents the phase portrait reconstructed from the time series, revealing a complex, ergodic, and bounded topology characteristic of high-dimensional chaotic attractors. To quantify the dynamic complexity, we compute the correlation dimension (CD) using the Grassberger-Procaccia algorithm[38]. As shown in Fig. 2g, the correlation integral $C_D(r)$ scales linearly with radius $r$ in the log-log domain at small $r$, with the slope converging to CD ≈ 8.12 (Fig. 2h), significantly higher than that of conventional optical feedback chaos[16]. More details about the characterization method and parameter influences are provided in Supplementary Notes 3-6. These results confirm that the "chaos-on-comb" architecture introduces enhanced nonlinear dynamics, enabling generation of high-complexity ultra-broadband chaotic signals.

Beyond generating a single ultra-broadband chaotic signal, our platform supports programmable, uncorrelated parallel chaotic outputs via a spectral waveshaper with independently tunable center wavelengths and filter bandwidths. Figure 3a-e demonstrates this programmability through four representative channels. Time-domain traces (Fig. 3a) exhibit strong stochastic fluctuations, while their radio-frequency spectra (Fig. 3b) show no relaxation oscillation peaks. Correlation integral slopes (Fig. 3c) confirm high dynamical complexity across channels. Figure 3d plots the effective bandwidth as a function of filter bandwidth for four wavelengths, revealing a synchronous increase in effective bandwidth with broader filters across all channels. These results validate the capability of our "chaos-on-comb" architecture to flexibly tailor chaotic channels, each with independently defined spectral properties (More details are provided in Supplementary Note 7). Inter-channel correlation is a critical metric for evaluating parallel chaotic signals. To assess this, we generate 16 parallel channels with 40 GHz filter bandwidth and 50 GHz channel spacing (the characterization of these optical channels and the influence of channel spacing are shown in Supplementary Notes 8 and 9). The resulting correlation matrix is shown in Fig. 3e. While slightly elevated correlations are observed between adjacent and diagonal channels, all inter-channel correlation coefficients remain below 0.4, satisfying the requirements for parallel operation.

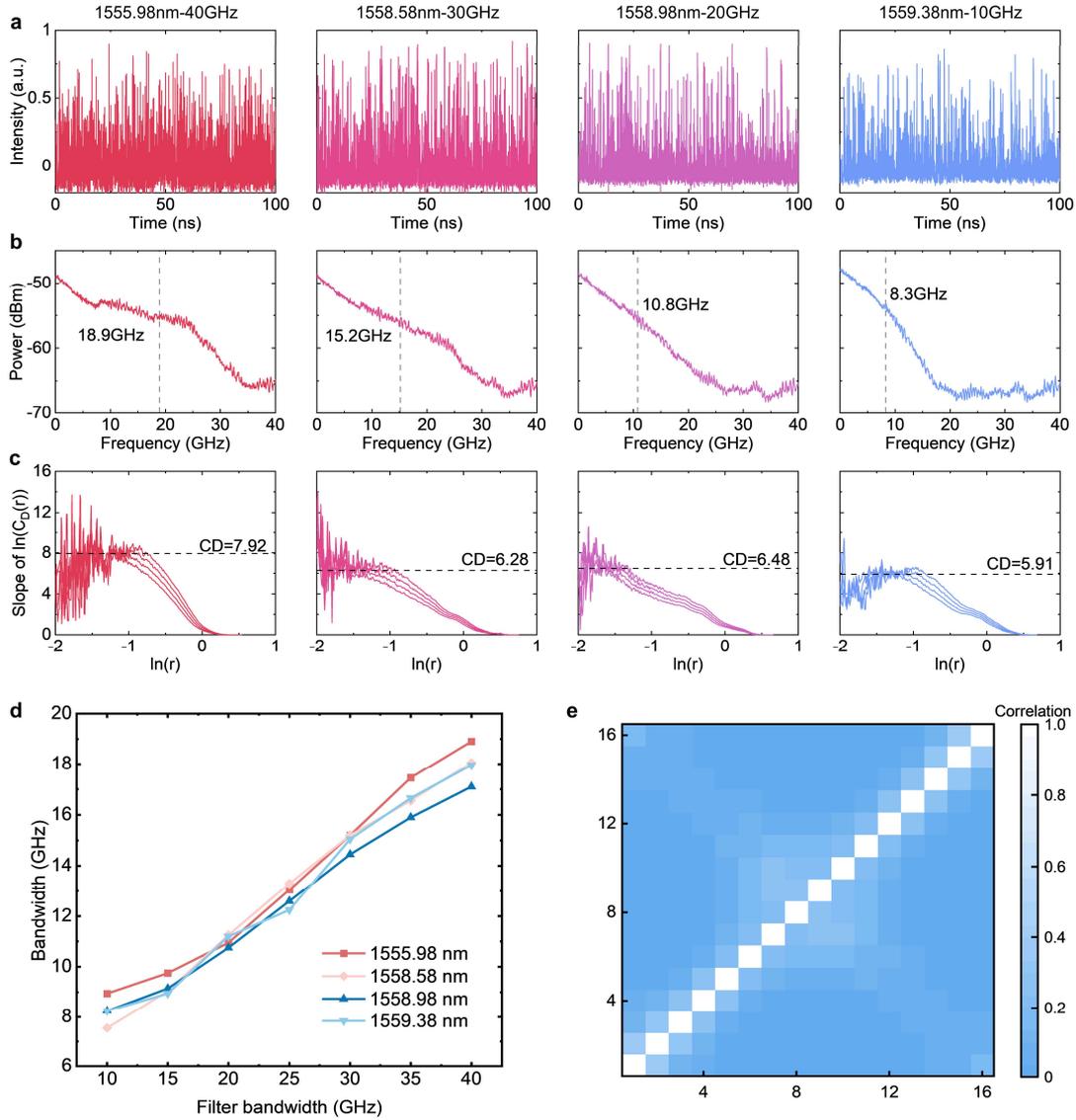

**Fig. 3 Characteristics of programmable parallel chaotic outputs. a-c**, Time series (**a**), radio-frequency spectra (**b**), and correlation integral slopes (**c**) of four representative chaotic outputs with varying center wavelengths and filtering bandwidths. **d**, Effective bandwidth as a function of filter bandwidth for four different wavelengths. **e**, Correlation matrix of 16 chaotic channels generated with 40 GHz filter bandwidth and 50 GHz spacing.

**Broadband terahertz noise source generation**

To demonstrate the ultra-broadband nature of the proposed single-channel chaotic scheme, we apply it to broadband terahertz noise source generation. As illustrated in Fig. 4a, the ultra-broadband chaotic signal is combined with a CW laser via a fiber coupler, with the optical power ratio precisely adjusted to optimize noise characteristics. Spectral flatness is critically governed by the frequency detuning between the chaotic signal and the CW light; in our system, this detuning is set to 320 GHz, yielding maximum flatness across the 0-500 GHz band (See more details in Supplementary Note 10). The combined signal is subsequently amplified by an erbium-doped fiber amplifier (EDFA) and converted into an electrical noise source using a uni-traveling-carrier photodiode (UTC-PD).

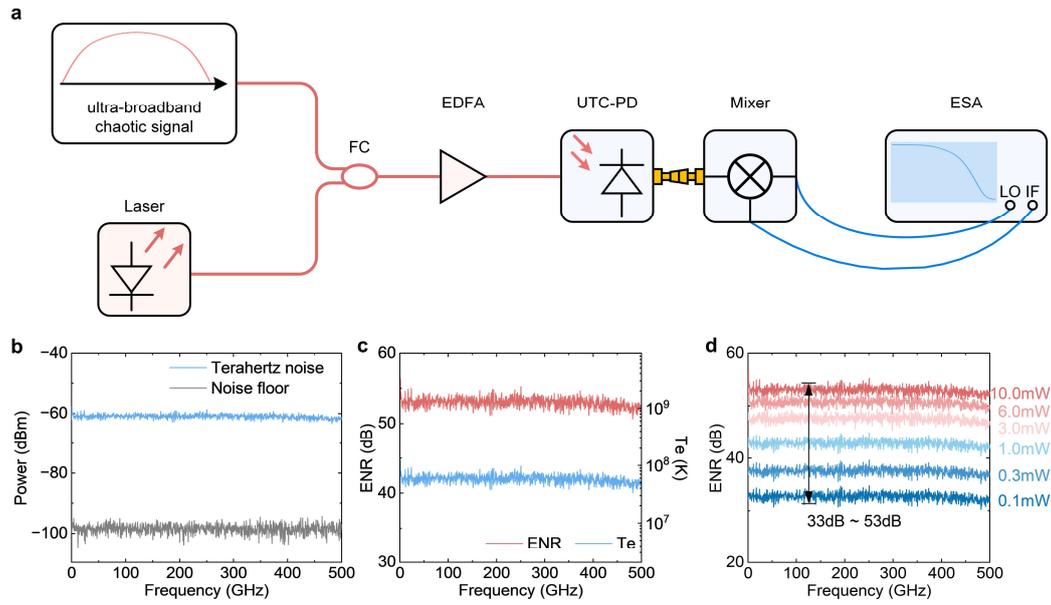

**Fig. 4 Broadband terahertz noise source enabled by ultra-broadband photonic chaos. a**, Schematic of the simulation setup for noise source generation. FC, fiber coupler; UTC-PD, uni-traveling-carrier photodiode; ESA, electrical spectrum analyzer. **b**, Measured power spectral density of the generated terahertz noise (blue) compared with the system noise floor (gray). **c**, Calculated ENR and noise equivalent temperature across the 0-500 GHz band. **d**, ENR spectra at different input optical powers.

Figure 4b shows the measured terahertz noise spectrum (blue curve) and the system noise floor (gray curve). The generated noise exhibits excellent flatness across the full 0-500 GHz bandwidth, with a power spectral density approximately 40 dB above the system baseline, demonstrating high spectral uniformity and strong excess noise performance. To quantify this, we calculate the excess noise ratio (ENR) and corresponding noise equivalent temperature (NET) (see Supplementary Note 11), as shown in Fig. 4c. Under optimized operating conditions, the source achieves an ENR of $52.99 \pm 2.85$ dB and a mean NET of $5.85 \times 10^7$ K across the 0-500 GHz range, surpassing most reported noise sources in both bandwidth and thermal equivalence. The

ENR of the generated noise source is readily tunable by adjusting the output power of the EDFA. Figure 4d plots ENR as a function of frequency under varying optical powers. The system achieves a continuously adjustable ENR from 32 to 53 dB while preserving excellent spectral flatness.

**Decision-making using programmable ultra-broadband photonic chaos**

The MAB problem represents a fundamental challenge in reinforcement learning, centered on achieving an optimal trade-off between exploration and exploitation under unknown reward distributions[39]. Photonic decision-making systems leveraging optical chaos offer a promising path toward enhanced computational efficiency and scalability[40-42], benefiting from the inherent randomness, high-dimensional dynamics, and multi-channel parallelism of chaotic signals. In this section, we present a scalable photonic computation acceleration framework based on our programmable ultra-broadband chaos-on-comb platform. Figure 5a illustrates the scheme of the proposed photonic computation acceleration system. Each channel $I_i(t)$ ($i = 1…N$) corresponds to a slot machine with an unknown reward probability $P_i$, and the objective is to find the machine with the highest $P_i$. Within the decision maker, each chaotic signal $I_i(t)$ is digitalized to 8 bits with 4 least significant bits retained, followed by a time-delayed exclusive-OR (XOR) operation to extract a high-entropy random sequence $A_i(t)$. The decision signal $D_i(t)$ is computed as:

$$D_i(t) = A_i(t) + k \cdot B_i(t), \tag{1}$$

where $k$ is a tunable bias coefficient and $B_i(t)$ is a time-varying bias term that reflects accumulated reward information. The channel with the highest $D_i(t)$ is selected as the optimal arm. The bias values $B_i(t)$ are iteratively updated using the tug-of-war method[40] to dynamically balance exploration and exploitation (see "Methods" for details).

In the experiment, the programmable photonic chaos platform is configured via a waveshaper to generate 16 uncorrelated parallel chaotic channels (as characterized in Fig. 3e), which are sequentially detected to solve a 16-armed bandit problem. The reward probabilities are set as: $P_1 = 0.7$, $P_2 = 0.5$, $P_3 = 0.9$, $P_4 = 0.1$, ..., $P_{2j-1} = 0.7$, $P_{2j} = 0.5$, where slot machine 3 has the highest reward probability. Figure 5b and c shows

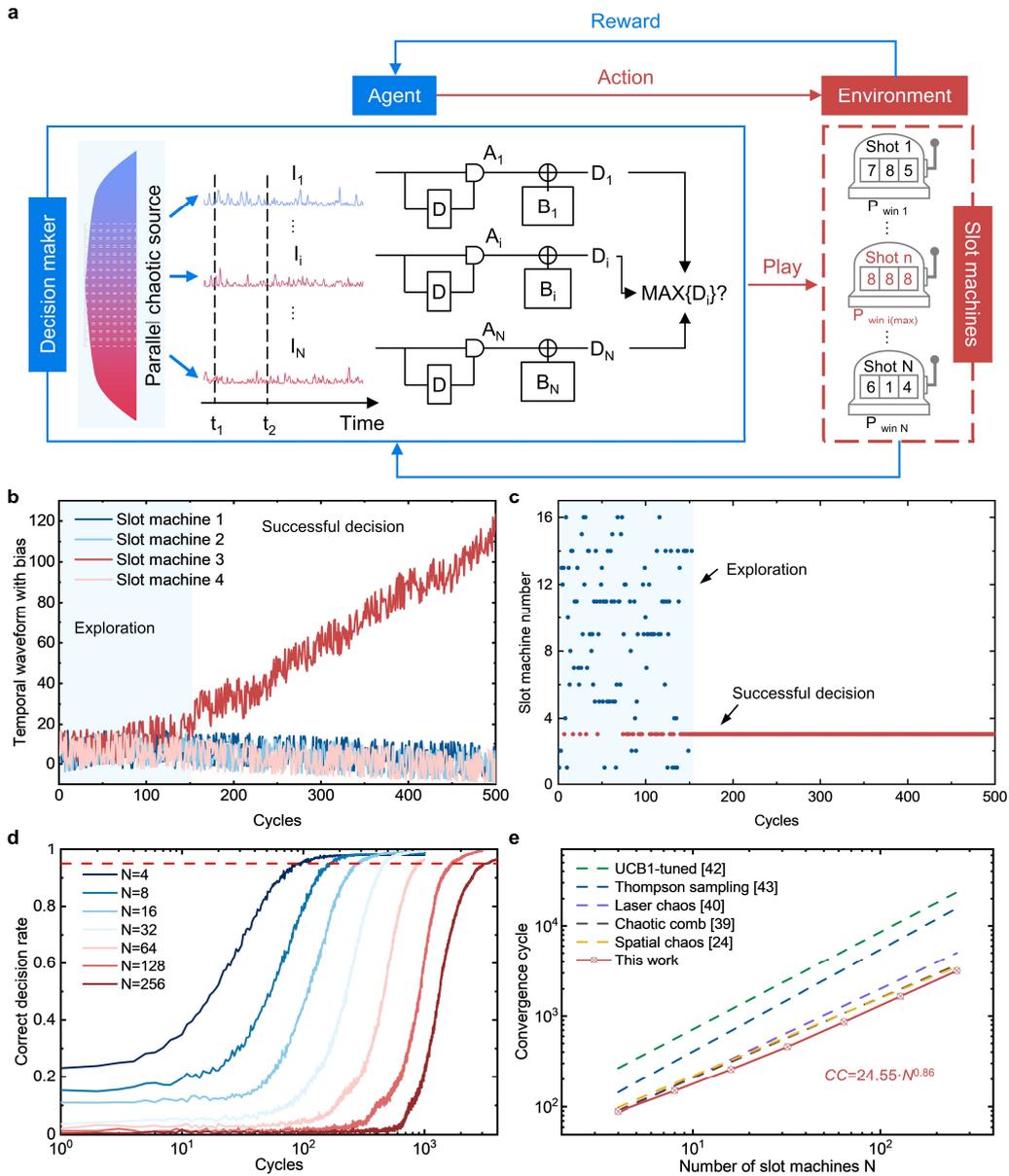

**Fig. 5 Solving the multi-armed bandit problem using programmable parallel photonic chaos.** **a**, architecture of the photonic decision-making accelerator leveraging 16-channel uncorrelated chaotic signals generated by the programmable photonic chaos platform. **b**, Temporal waveforms of biased chaotic decision signals for slot machines 1~4. **c**, Example decision sequence in a 16-armed bandit problem, where slot machine 3 (red dots) with the highest reward probability is consistently selected after convergence. **d**, Correct decision rate as a function of the number of slot machine arms. **e**, Comparison of convergence cycles versus arm count between our method and prior approaches.

the evolution of the decision-making process. In Fig. 5b, the biased chaotic signals $D_i(t)$

show comparable amplitudes across all arms during the initial trials (only slots 1-4 are displayed for clarity), reflecting the exploration process. Subsequently, the amplitude of slot machine 3 becomes dominant while others remain suppressed, indicating convergence to the optimal decision. Correspondingly, Fig. 5c traces the selected arm index over successive decisions. While the system explores multiple arms in early trials, it consistently selects slot machine 3 after approximately the 150$^{th}$ iteration, demonstrating stable identification of the optimal arm.

To evaluate the decision-making performance, we define the correct decision rate (CDR) as CDR = $T_{hit}/T$, where $T_{hit}$ is the number of successful selections in each decision cycle over $T$ = 1000 independent trials. A CDR above 95% is regarded as a successful decision, and the first cycle at which this threshold is reached is defined as the convergence cycle (*CC*). Figure 5d presents the accuracy and scalability of the photonic decision-making system. For $N > 16$, time-multiplexed recordings of the 16 available channels are used to synthesize higher-dimensional input. As $N$ increases to 256, *CC* increases accordingly while maintaining CDR > 95%, demonstrating robust scalability to large-scale MAB problems. Figure 5e benchmarks our system against existing approaches for solving MAB problems[24,40,41,43,44]. Power-law fitting yields the empirical relation $CC=24.55\cdot N^{0.86}$, indicating a favourable sublinear scaling exponent. Compared to prior methods, our approach achieves the fastest convergence with the lowest scaling exponent, making it well-suited for large-scale decision tasks. Moreover, by reducing filter bandwidth and channel spacing, more chaotic channels can be accommodated, enabling higher decision throughput without compromising scalability.

**Discussion**

In conclusion, we demonstrate a programmable "chaos-on-comb" platform that generates ultra-broadband photonic chaos and validates its performance across two applications. By driving a standard electro-optic frequency comb with a broadband chaotic microwave signal, the system produces overlapping chaotic comb lines, achieving the highest single-channel chaotic bandwidth reported to date. Leveraging an optical waveshaper, the platform also enables independent control over the center wavelength and bandwidth of each output, supporting low-correlation parallel chaos generation with up to 50 programmable channels. Building on this system, we demonstrate high-performance terahertz noise generation and photonic decision acceleration for MAB problems, highlighting its strong potential for advanced technological applications. The platform addresses two longstanding limitations in optical chaos: it surpasses the relaxation-oscillation-imposed bandwidth ceiling, achieving an order-of-magnitude enhancement in effective bandwidth; and it resolves the trade-off between channel count and per-channel bandwidth, delivering >15 GHz effective bandwidth across 16 parallel, low-correlation channels. These advances may open new directions for deploying optical chaos in high-speed, multi-channel information processing.

The bandwidth of the ultra-broadband chaotic optical signal can be further extended via two main approaches. First, increasing the bandwidth of the chaotic seed enhances the spectral width of each comb tooth. In this work, the chaotic seed, generated via external optical feedback, has a bandwidth of ~9 GHz, which limits the effective spectral width per tooth. To ensure complete sideband aliasing, the comb repetition rate $\varDelta\omega$ is set to 32.5 GHz (See Supplementary Note 4). By adopting advanced techniques such as nonlinear spectral broadening to expand the seed bandwidth, $\varDelta\omega$ can be increased accordingly, thereby yielding a broader chaotic spectrum. Second, cascading additional electro-optic modulators boosts spectral broadening through high-order sideband generation. The bandwidth gain in this architecture is primarily driven by phase modulation, and employing modulators with lower half-wave voltage or cascading multiple phase modulators enables further extension. In our experiment, two cascaded phase modulators yield a >700 GHz optical bandwidth, while a triple-cascade configuration is anticipated to generate a flat chaotic spectrum exceeding 1 THz.

It is worth emphasizing that the current experimental implementation is constructed using discrete optical components. However, all core functional modules, including chaotic seed signal[45,46], electro-optic modulators[47,48], and auxiliary optical components[49,50], have mature pathways toward photonic integration. With advances in on-chip integration and photonic integrated circuit technology[51,52], the entire chaos-on-comb architecture could be miniaturized onto a chip-scale platform. This transition would substantially reduce system size, power consumption, and cost, paving the way for scalable and practical deployment in real-world applications.

Beyond demonstrated applications in noise generation and photonic decision-making, the unique capabilities of the proposed chaos-on-comb platform offer transformative potential across a range of frontier domains. In LiDAR, the ultrabroadband spectrum can significantly enhance ranging resolution. In secure communications, the high-entropy chaotic carriers support elevated data rates and improved encryption robustness. Moreover, the platform is well-suited for high-throughput information processing tasks, including ultrafast random number generation and photonic probabilistic computing. These application scenarios highlight the platform's versatility and establish it as a practical and scalable solution for pushing the performance boundaries of optical chaos technologies.

## Methods

**Experimental setup**

The experimental configuration is depicted in Fig. 2a. A chaotic seed signal is generated by a distributed feedback (DFB) laser at 1549.66 nm under optical feedback, with 30% of its output reflected via a 3:7 fiber coupler and mirror. A polarization controller and variable optical attenuator (VOA) are used to manage the feedback state. The chaotic seed signal is combined with a 1549.92 nm CW reference from a tunable laser (Keysight, N7711A) using a 90° optical hybrid and heterodyne-detected by a balanced photodetector (BPD), producing a 32.5 GHz chaotic microwave signal through frequency down-conversion. After RF amplification (MLA-100k50G-N5.0G25P), the chaotic microwave signal is split into three branches via a power divider (Qotana, DBPD041800400A) to synchronously drive a cascaded electro-optic modulation system that consists of an intensity modulator (Conquer, Photonics KG-AM-15-40G-PP-FA) and two phase modulators (Eospace, PM-5VEK-40-PFA-UV). Phase shifters ensure optimal timing, while the drive power for the PMs is amplified to ~ 33 dBm to maximize the PM modulation depth (3.1 $\pi$) and spectral broadening. A CW optical carrier at 1558.98 nm with 13 dBm power sequentially passes through the

IM and PMs. The modulated signal then propagates through 10 km of SMF for dispersion-induced phase-to-intensity conversion, with an EDFA compensating loss. The resulting ultra-broadband chaotic signal is spectrally sliced using a programmable optical filter (Coherent, WaveShaper 4000A) to enable flexible parallel chaotic outputs. The optical output is analyzed via an optical spectrum analyzer (Yokogawa, AQ6370D), or detected by a 40 GHz photodetector (OVLINK, PD-1000) and measured using a real-time digital phosphor oscilloscope (Keysight, DSA V204A, 20 GHz, 80 GSa/s) and an electric spectrum analyzer (Ceyear, 4052H, 50 GHz).

**Decision-Making Method**

This work uses a photonic decision-making strategy based on the tug-of-war method. In our experiment, time series from 16 chaotic channels $I_i(t)$ are sequentially acquired and filtered to extract random sequences $A_i(t)$ via digital post-processing. The biased decision signal $D_i(t)$ is calculated according to Eq.(1), where $B_i(t)$ is a dynamical bias term computed as follows:

$$B_i(t) = Q_i(t) - \frac{1}{N-1} \sum_{i' \neq i}^{N} Q_{i'}(t) \qquad (2)$$

$$Q_i(t) = \alpha * W_i - \beta * L_i \qquad (3)$$

$$\alpha = 2 - \hat{P}_{top1} - \hat{P}_{top2} \qquad (4)$$

$$\beta = \hat{P}_{top1} + \hat{P}_{top2} \qquad (5)$$

$$\hat{P}_i = \frac{W_i}{T_i} \qquad (6)$$

where $Q_i(t)$ represents the evaluation value of slot machine $i$ at cycle $t$, and $N$ is the total number of slot machines. $T_i$ is the total number of times slot machine $i$ has been selected, and $W_i$ and $L_i$ are the respective counts of wins and losses. $\hat{P}_i$ denotes the estimated win probability of slot machine $i$, and $\hat{P}_{top1}$ and $\hat{P}_{top2}$ are the highest and second-highest estimated probabilities among all slots. The arm with the maximum $D_i(t)$ is selected. The bias coefficient $k$ regulates the trade-off between exploration and exploitation, and its value is optimized for different slot machine counts to maximize decision-making performance.

**References**

1. May, R. M. Simple mathematical models with very complicated dynamics. Nature 261, 459-467 (1976).
2. Soriano, M. C., García-Ojalvo, J., Mirasso, C. R. & Fischer, I. Complex photonics: Dynamics and applications of delay-coupled semiconductors lasers. Rev. Mod. Phys. 85, 421-470 (2013).
3. Zhao, A., Jiang, N., Liu, S., Zhang, Y. & Qiu, K. Physical layer encryption for WDM optical communication systems using private chaotic phase scrambling. J. Lightwave Technol. 39, 2288-2295 (2021).
4. Zhang, Y. et al. Simultaneously enhancing capacity and security in free-space optical chaotic communication utilizing orbital angular momentum. Photonics Res. 11 (2023).
5. Feng, J. et al. 256 Gbit/s chaotic optical communication over 1600km using an AI-based optoelectronic oscillator model. J. Lightwave Technol., 1-10 (2024).
6. Didier, P. et al. Data encryption with chaotic light in the long wavelength infrared atmospheric window. Optica 11 (2024).
7. Asmussen, S. & Glynn, P. W. Stochastic Simulation: Algorithms and Analysis. Vol. 57 (Springer, 2007).
8. Xiang, S. et al. 2.24-Tb/s physical random bit generation with minimal post-processing based on chaotic semiconductor lasers network. J. Lightwave Technol. 37, 3987-3993 (2019).
9. Zhao, L. et al. 126 Tbits/s massive parallel physical random bits generator with broadband chaos of integrated AlGaAs micro-resonator. Laser Photonics Rev. (2025).
10. Bruckerhoff-Pluckelmann, F. et al. Probabilistic photonic computing with chaotic light. Nat. Commun. 15, 10445 (2024).
11. Liu, L. et al. Universal millimeter-wave noise source based on a multi-mode chaotic laser. Opt. Laser Technol. 181 (2025).
12. Liu, W. et al. Broadband and flat millimeter-wave noise source based on the heterodyne of two Fabry–Perot lasers. Opt. Lett. 47, 541-544 (2022).
13. Xu, Z. et al. High-resolution radar ranging based on the ultra-wideband chaotic optoelectronic oscillator. Opt. Express 31, 22594-22602 (2023).
14. Lukashchuk, A., Riemensberger, J., Tusnin, A., Liu, J. & Kippenberg, T. J. Chaotic microcomb-based parallel ranging. Nat. Photonics 17, 814-821 (2023).
15. Sun, Y. et al. Multi-color pulsed chaos enables single-pixel parallel laser ranging. PhotoniX 6 (2025).
16. Sciamanna, M. & Shore, K. A. Physics and applications of laser diode chaos. Nat. Photonics 9, 151-162 (2015).
17. Ohtsubo, J. in Semiconductor Lasers: Stability, Instability and Chaos 33-82 (Springer, 2017).
18. Yang, Q. et al. Generation of a broadband chaotic laser by active optical feedback loop combined with a high nonlinear fiber. Opt. Lett. 45, 1750-1753 (2020).
19. Li, P. et al. Observation of flat chaos generation using an optical feedback multi-mode laser with a band-pass filter. Opt. Express 27, 17859-17867 (2019).
20. Han, Y. et al. Generation of multi-channel chaotic signals with time delay signature concealment and ultrafast photonic decision making based on a globally-coupled semiconductor laser network. Photonics Res. 8 (2020).

21. Liu, S. et al. Generation of multiple low-correlation chaos signals using asymmetric coupling semiconductor lasers networks. Opt. Laser Technol. 155 (2022).
22. Huang, Y., Zhou, P., Lau, K. & Li, N. Parallel wideband chaos generation system for advancing high-throughput information processing based on an array of four distributed feedback lasers. ACS Photonics 11, 5012-5021 (2024).
23. Chen, R. et al. Breaking the temporal and frequency congestion of LiDAR by parallel chaos. Nat. Photonics 17, 306-314 (2023).
24. Shen, B. et al. Harnessing microcomb-based parallel chaos for random number generation and optical decision making. Nat. Commun. 14, 4590 (2023).
25. Cuevas, J., Iwami, R., Uchida, A., Minoshima, K. & Kuse, N. Solving multi-armed bandit problems using a chaotic microresonator comb. APL Photonics 9 (2024).
26. Li, J. C., Xiao, J. L., Yang, Y. D., Chen, Y. L. & Huang, Y. Z. Random bit generation based on a self-chaotic microlaser with enhanced chaotic bandwidth. Nanophotonics 12, 4109-4116 (2023).
27. Bouchez, G., Uy, C. H., Macias, B., Wolfersberger, D. & Sciamanna, M. Wideband chaos from a laser diode with phase-conjugate feedback. Opt. Lett. 44, 975-978 (2019).
28. Wang, Y. et al. Generation of laser chaos with wide-band flat power spectrum in a circular-side hexagonal resonator microlaser with optical feedback. Opt. Express 28, 18507-18515 (2020).
29. Zhao, A. et al. Parallel generation of low-correlation wideband complex chaotic signals using CW laser and external-cavity laser with self-phase-modulated injection. Opto-Electronic Advances 5, 200026-200026 (2022).
30. Cai, Q. et al. Tbps parallel random number generation based on a single quarter-wavelength-shifted DFB laser. Opt. Laser Technol. 162 (2023).
31. Zhao, A. et al. Semiconductor laser-based multi-channel wideband chaos generation using optoelectronic hybrid feedback and parallel filtering. J. Lightwave Technol. 40, 751-761 (2022).
32. Zhang, Q. et al. Multi-channel broadband optical chaos generation assisted by phase modulation and CFBG feedback. Opt. Express 32, 20471-20482 (2024).
33. Li, Q., Jia, Z., Wang, A. & Wang, Y. Parallel generation of multi-channel broadband chaos by a long-cavity FP laser with optical feedback. Opt. Lett. 49, 7126-7129 (2024).
34. Hu, C.-X. et al. Simultaneous generation of multi-channel broadband chaotic signals based on two unidirectionally coupled WRC-FPLDs. IEEE Photonics J. 12, 1-8 (2020).
35. Li, S. S., Li, X. Z. & Chan, S. C. Chaotic time-delay signature suppression with bandwidth broadening by fiber propagation. Opt. Lett. 43, 4751-4754 (2018).
36. VPIphotonics: Simulation software and design services. Online: https://www.vpiphotonics.com/index.php (2025).
37. Lin, F.-Y., Chao, Y.-K. & Wu, T.-C. Effective bandwidths of broadband chaotic signals. IEEE J. Quantum Electron. 48, 1010-1014 (2012).
38. Grassberger, P. & Procaccia, I. Estimation of the Kolmogorov entropy from a chaotic signal. Phys. Rev. A 28, 2591-2593 (1983).
39. Matsuo, Y. et al. Deep learning, reinforcement learning, and world models. Neural. Netw. 152, 267-275 (2022).
40. Morijiri, K. et al. Parallel photonic accelerator for decision making using optical

spatiotemporal chaos. Optica 10 (2023).
41. Morijiri, K., Mihana, T., Kanno, K., Naruse, M. & Uchida, A. Decision making for large-scale multi-armed bandit problems using bias control of chaotic temporal waveforms in semiconductor lasers. Sci. Rep. 12, 8073 (2022).
42. Xu, Z. et al. Harnessing nonlinear optoelectronic oscillator for speeding up reinforcement learning. PhotoniX 6 (2025).
43. Auer, P., Cesa-Bianchi, N. & Fischer, P. Finite-time analysis of the multiarmed bandit problem. Mach. Learn. 47, 235-256 (2002).
44. Thompson, W. R. On the likelihood that one unknown probability exceeds another in view of the evidence of two samples. Biometrika 25, 285-294 (1933).
45. Chai, M. et al. Wavelength-tunable monolithically integrated chaotic semiconductor laser. J. Lightwave Technol. 40, 5952-5957 (2022).
46. Wang, Y. et al. Hybrid integrated optical chaos circuits with optoelectronic feedback. Opt. Express 32 (2024).
47. Zhang, K. et al. A power-efficient integrated lithium niobate electro-optic comb generator. Commun. Phys. 6 (2023).
48. Qi, Y., Yue, G., Hao, T. & Li, Y. Strong-confinement low-index-rib-loaded waveguide structure for etchless thin-film integrated photonics. Opto-Electronic Advances 8 (2025).
49. Zhou, J. et al. On-chip integrated waveguide amplifiers on erbium-doped thin-film lithium niobate on insulator. Laser Photonics Rev. 15 (2021).
50. Xie, Y. et al. Low-loss chip-scale programmable silicon photonic processor. Opto-Electronic Advances 6, 220030-220030 (2023).
51. Huang, H. et al. High-intensity spatial-mode steerable frequency up-converter toward on-chip integration. Opto-Electronic Science 3, 230036-230036 (2024).
52. Di Falco, A., Mazzone, V., Cruz, A. & Fratalocchi, A. Perfect secrecy cryptography via mixing of chaotic waves in irreversible time-varying silicon chips. Nat. Commun. 10, 5827 (2019).

**Acknowledgements**

This work was supported by the National Natural Science Foundation of China (NSFC) (Grants Nos. U24A6010, 62575286, 62401540) and China Postdoctoral Science Foundation (Grant No. 2024M753238).

**Author contributions**

Y.Q.Z., M.F.X. conceived the idea and planned the research. S.Y.S., M.J.Z., X.L.Y., and Y.X.Y. performed the ultra-broadband optical chaos simulations. S.Y.S., M.J.Z., and X.L.Y. designed and conducted experiments. Y.Q.Z., S.Y.S., and Q.C. contributed to the data analysis. S.Y.S., Y.Q.Z., and M.F.X. wrote the manuscript. Y.H.G., and M.B.P. presented suggestions for improving the quality of this work. M.B.P., and X.G.L. supervised the project. All authors discussed and analyzed the data and results.

**Competing interests**

**Data availability**

All the data used to generate the plots and support the findings reported in this study are available from the corresponding authors upon request.